\documentclass[12pt]{article}

\newcommand{\be}{\begin{equation}}
\newcommand{\ee}{\end{equation}}
\newcommand{\bi}[1]{\vspace{-3mm} \bibitem{#1}}

\newcommand{\et}{\rm \\ }

\newcommand{\bp}{{\bf Proof.} }

\usepackage{epsfig,amsmath,amssymb,graphics,graphicx} 

\usepackage{graphics,graphicx}
\usepackage{amssymb,amsmath}
\usepackage{amscd}
\usepackage{pb-diagram}
\usepackage{hyperref}
\usepackage{amsfonts}
\usepackage{epsfig}

\begin{document}

\begin{center}

{\it Central European Journal of Physics. Vol.11. No.11. (2013) 1580-1588} \\

\vskip 3mm

{\bf \large Lattice Model with Power-Law Spatial Dispersion  \\
\vskip 3mm
for Fractional Elasticity} \\

\vskip 7mm
{\bf \large Vasily E. Tarasov} \\
\vskip 3mm

{\it Skobeltsyn Institute of Nuclear Physics,\\ 
Lomonosov Moscow State University, Moscow 119991, Russia} \\
{E-mail: tarasov@theory.sinp.msu.ru} \\

\begin{abstract}
A lattice model with a spatial dispersion corresponding to a power-law type is suggested.
This model serves as a microscopic model for elastic continuum
with power-law non-locality.
We prove that the continuous limit maps of the equations for the lattice 
with the power-law spatial dispersion into the continuum equations 
with fractional generalizations of the Laplacian operators.
The suggested continuum equations, which are obtained from the lattice model, 
are fractional generalizations of the integral and gradient elasticity models.
These equations of fractional elasticity are solved for
two special static cases:  fractional integral elasticity and fractional gradient elasticity.
\end{abstract}

\end{center}


\vskip 3mm

\section{Introduction}

The integration and differentiation of any arbitrary order 
have a long history since 1695 \cite{Ross} - \cite{SKM2}. 
The theory of equations with derivatives and integrals of non-integer orders \cite{SKM1,SKM2,KST} 
is powerful tool to describe the behavior of media and systems with 
spatial non-locality and memory of a power-law type. 
Fractional differential equations have a vast applications in 
physics and mechanics and form an area called fractional dynamics \cite{CM} - \cite{IJMPB2013}.
The fractional calculus allows us to formulate 
a fractional generalization of non-local elasticity models in two forms:
the fractional differential (gradient) elasticity models (weak power-law non-locality) and 
the fractional integral models (strong power-law non-locality).
Fractional integral elasticity models are considered in 
\cite{Laz} - \cite{CPZ-3}.

Lattice models are very important in elasticity theory 
(see for example \cite{Kunin} -\cite{CT2012}).
In this paper we suggest a lattice model with a power-law spatial dispersion as
a microscopic model of elastic continuum with weak and strong power-law non-locality. 
For this purpose, we consider the models of lattices and 
their corresponding continuous limits by using the methods suggested in \cite{JPA2006,JMP2006} 
(see also \cite{Chaos2006,LZ,CNSNS2006}).
In \cite{JPA2006,JMP2006} we prove that the continuum equations with fractional Laplacian
in the Riesz form \cite{SKM1,SKM2,KST} can be directly derived from lattice models with 
different types of long-range interactions. 
In this paper, we show how the continuous limit for the lattice with 
a power-law spatial dispersion gives the corresponding 
continuum equation of the fractional integral and gradient elasticity.
The continuum equations of fractional elasticity, which are derived from the lattice model, 
are solved for two static cases: fractional integral elasticity and fractional gradient elasticity.

\section{Equations for Displacement of Lattice Particles}

Let us consider a lattice model  
where all particles are displaced in one direction.
We assume that the displacement of particles from its equilibrium position 
is determined by a scalar field. 
It allows us to describe the properties of the lattice and its continuum limit 
by using simple equations. 

The equations of motion for a one-dimensional lattice system of 
interacting particles have the form
\be \label{Main_Eq_2}
M \, \frac{\partial^2 u_n(t)}{\partial t^2} = g 
\sum_{\substack{m=-\infty \\ m \ne n}}^{+\infty} \; K(n,m) \; \Bigl( u_n(t)-u_m (t) \Bigr) 
 + F(n) ,
\ee
where $u_n(t)= u(n,t)$ is the displacement of $n$ particle from its equilibrium position,   
$g$ is the coupling constant for interparticle interactions in the lattice, 
the terms $F(n)$ characterize an interaction of the particles   
with the external on-site force.
For simplicity, we assume that all particles have the same mass $M$. 
The elements $K(n,m)$ of equation (\ref{Main_Eq_2})
describe the interparticle interaction in the lattice. 
For an unbounded homogeneous lattice, due to its homogeneity,   
$K(n,m)$ has the form $K( n, m ) = K(n -  m)$.
 Equations of motion (\ref{Main_Eq_2}) is invariant 
with respect to its displacement of the lattice as a whole, 
provided external forces are absent.
It should be noted that the noninvariant terms lead to divergences 
in the continuous limit \cite{TarasovSpringer}.

\section{Transform Operation for Lattice Models}

In order to define the operation that transforms
the lattice equations for $u_n(t)$
into the continuum equation for a scalar field $u(x,t)$, 
we use the methods suggested in \cite{JPA2006,JMP2006}:
We consider $u_n(t)$ as Fourier series coefficients
of some function $\hat{u}(k,t)$ on $[-k_0/2, k_0/2]$,
then we use the continuous limit $k_0 \to \infty$ to obtain $\tilde{u}(k,t)$,
and finally we apply the inverse Fourier integral transformation to obtain $u(x,t)$. 
Diagrammatically this can be written in the following form.

\be 
\begin{CD}
u_n(t) @> {{\cal F}_{\Delta} } >> \hat{u}(k,t) @>{ \operatorname{Lim} }>> \tilde{u}(k,t) 
@>{ {\cal F}^{-1}  }>> u(x,t)
\end{CD} \ee

We performed similar transformations for differential equations 
to map the lattice equation into an equation for the elastic continuum.
We can represent these sets of operations in the form of the following diagrams.

\be \label{Di-2}
\begin{CD}
\boxed{Equation \ for \ u_n(t)} @>{ From \ Lattice \ to \ Continuum}>> \boxed{Equation \ for \ u(x,t)} \\
@V{ Fourier \ series \ transform }VV @A{ Inverse \ Fourier \ integral \ transform }AA \\
\boxed{Equation \ for \ \hat{u}(k,t)} @>>{ Limit \ \Delta x \to 0  }> \boxed{Equation \ for \ \tilde{u}(k,t)} 
\end{CD}
\ee

Therefore the transformation operation that maps our lattice model into a continuum model 
is a sequence of the following three actions (for details see \cite{JPA2006,JMP2006}):
\begin{enumerate} 
\item
The Fourier series transform 
${\cal F}_{\Delta}: \quad u_n(t) \to {\cal F}_{\Delta}\{ u_n(t)\}= \hat{u}(k,t) $ 
that is defined by
\be \label{ukt}
\hat{u}(k,t) = \sum_{n=-\infty}^{+\infty} \; u_n(t) \; e^{-i k x_n} =
{\cal F}_{\Delta} \{u_n(t)\} ,
\ee
\be \label{un} 
u_n(t) = \frac{1}{k_0} \int_{-k_0/2}^{+k_0/2} dk \ \hat{u}(k,t) \; e^{i k x_n}= 
{\cal F}^{-1}_{\Delta} \{ \hat{u}(k,t) \} ,
\ee
where $x_n = n \Delta x$, and $\Delta x=2\pi/k_0$ is the inter-particle distance.
For simplicity we assume that all 
lattice particles have the same inter-particle distance $\Delta x$.

\item 
The passage to the limit $\Delta x \to 0$ ($k_0 \to \infty$) denoted by
$\operatorname{Lim}: \quad \hat{u}(k,t) \to \operatorname{Lim} \{\hat{u}(k,t)\}= \tilde{u}(k,t)$. 
The function $\tilde{u}(k,t)$ can be derived from $\hat{u}(k,t)$
in the limit $\Delta x \to 0$.
Note that $\tilde{u}(k,t)$ is a Fourier integral transform of the field $u(x,t)$,
and $\hat{u}(k,t)$ is a Fourier series transform of $u_n(t)$,
where we use 
\[ u_n(t) = \frac{2 \pi}{k_0} u(x_n,t) \] 
considering $x_n=n\Delta x= 2 \pi n /k_0 \to x$.

\item 
The inverse Fourier integral transform 
${\cal F}^{-1}: \quad \tilde{u}(k,t) \to {\cal F}^{-1} \{ \tilde{u}(k,t)\}=u(x,t)$
is defined by 
\be \label{ukt2} 
\tilde{u}(k,t)=\int^{+\infty}_{-\infty} dx \ e^{-ikx} u(x,t) = 
{\cal F} \{ u(x,t) \}, 
\ee
\be \label{uxt}
u(x,t) =\frac{1}{2\pi} \int^{+\infty}_{-\infty} dk \ e^{ikx} \tilde{u}(k,t) =
 {\cal F}^{-1} \{ \tilde{u}(k,t) \} . 
\ee
\end{enumerate} 

Using the suggested notations we can represent diagram (\ref{Di-2}) in the following form,

\be \label{Di-3}
\begin{CD}
u_n(t) @>{ From \ Particle \ to \ Field}>> u(x,t) \\
@V{{ \cal F}_{\Delta} }VV @AA{ {\cal F}^{-1} }A \\
\hat{u}(k,t) @>>{ \operatorname{Lim \ \ \Delta x \to 0 } }> \tilde{u}(k,t) \ .
\end{CD} 
\ee

The combination of these three actions ${\cal F}^{-1}$, $\operatorname{Lim}$, and ${\cal F}_{\Delta}$ 
allows us to realize the transformation of lattice models 
into continuum models \cite{JPA2006,JMP2006}.  

Note that equations (\ref{ukt}) and (\ref{un}) 
in the limit $\Delta x \to 0$ ($k_0 \to \infty$) are 
used to obtain the Fourier integral transform equations  (\ref{ukt2}) and (\ref{uxt}),
where the sum is changed by the integral.

Let us give the statement that describes 
the Fourier series transform for the equations 
for displacement of lattice particles (\ref{Main_Eq_2}).
The Fourier series transform ${\cal F}_{\Delta}$ 
maps the lattice equations of motion 
\be \label{2-C1}
M \, \frac{\partial^2 u_n(t)}{\partial t^2}=g              
\sum^{+\infty }_{\substack{m=-\infty \\ m \not= n}}
K(n,m) \, \Bigl( u_n(t)-u_m(t) \Bigr) + F(n) ,
\ee
where $K(n,m)$ satisfies the conditions 
\be \label{2-Knm1}
K(n,m)=K(n-m)=K(m-n) ,  \quad
\sum^{\infty}_{n=1} |K(n)|^2 < \infty ,
\ee
into the continuum equation 
\be \label{2-20eq}
M \, \frac{\partial^2  \hat u(k,t)}{\partial t^2}=
g \Bigl( \hat{K} (0)- \hat{K} (k \Delta x) \Bigr) \, \hat u(k,t) 
+ {\cal F}_{\Delta} \{F(n)\} ,
\ee 
where   $\hat{u}(k,t)={\cal F}_{\Delta}\{ u_n(t)\}$,  \ 
$\hat{K} (k \Delta x)={\cal F}_{\Delta}\{ K(n)\}$, 
and ${\cal F}_{\Delta}$ is an operator notation for the Fourier
series transform.
Here we use the notation
\be \label{not}
\hat{K} (k \Delta x)=
\sum^{+\infty}_{\substack{n=-\infty \\ n\not=0}} 
e^{-ikn \Delta x} K(n) .
\ee
Using $K(-n)=K(n)$, the function (\ref{not}) can be represented by
\be \label{Kcos}
\hat{K} (k \Delta x) =\sum^{+\infty}_{n=1} K(n) 
\left( e^{-ikn\Delta x} +e^{ikn\Delta x} \right) = 
2\sum^{+\infty}_{n=1} K(n) \cos \left( k \Delta x \right) ,
\ee
and
\be \label{C8}
\hat{K} (0)=\sum^{+\infty}_{\substack{n=-\infty \\ n \not=0}}
K(n)=2\sum^{\infty}_{n=1} K(n) .
\ee
For details see \cite{JPA2006,JMP2006} and \cite{TarasovSpringer}. 

\section{Weak Spatial Dispersion of Power-Law Type}

In lattice models, dispersion is associated with different properties of the wave 
such as its frequency, wavelength, wave-number,  amplitude  and others.
Spatial dispersion is the dependence of the elastic waves on the wave vector.
In the model that is described by equation (\ref{Main_Eq_2})
the spatial dispersion means the dependence of the kernel $\hat{K} (|{\bf k}|)$
on the wave vector ${\bf k}$. 
This dependence is caused by non-local interactions in the elastic continuum.
The spatial dispersion leads to a non-local connection 
between the stress tensor $\sigma_{kl}$ and the strain tensor $\varepsilon_{kl}$. 
The tensor $\sigma_{kl}$ at any point ${\bf r}$ of the continuum is not uniquely defined by
the values of $\varepsilon_{kl}$ at this point. 
It also depends on the values of $\varepsilon_{kl}$ 
at neighboring points ${\bf r}^{\prime}$, located near the point ${\bf r}$.
Qualitatively describing the process in a lattice with spatial dispersion
implies that the fields of the elastic wave moves 
particles from their equilibrium positions 
at a given point ${\bf r}$, which causes an additional shift of the particles
in neighboring and more distant points ${\bf r}^{\prime}$ in some neighborhood region.
Therefore, the properties of the continuum, and hence the stress tensor field $\sigma_{kl}$ depend 
on the values of strain tensor field $\varepsilon_{kl}$ not only in a selected point, 
but also in some neighborhood region.
The size $R_0$ of the area of the mutual influence 
are usually of the order of the interparticle distance in the lattice. 
The wavelength, $\lambda$, of elastic waves is several orders larger than the size of this region, 
so for a region of size $R_0$ the values of the field of the elasticity wave do not change.
In other words, the wavelength $\lambda$ usually holds 
$k R_0 \sim R_0 / \lambda \ll 1$. 
In such a lattice the spatial dispersion is weak. 

To describe the dynamics of the lattice it is enough to know the dependence of 
the function $\hat{K} (|{\bf k}|)$ for small values of $k=|{\bf k}|$ only. Therefore
we can replace this function by the Taylor polynomial.
The weak spatial dispersion in the media with
a power-law type of non-locality cannot be described by
the usual Taylor approximation. 
The fractional Taylor series can be very useful for approximating 
non-integer power-law functions \cite{AP2013}. 
This is due to the fact that the usual Taylor series for the power-law function
has an infinite number of terms.
Using the fractional Taylor's formula we obtain a finite number of terms.

For an isotropic linear medium with the weak spatial dispersion 
the function  $\hat{K} (|{\bf k}|)$ can be represented in the form
\be \label{approx-1}
\hat{K} (|{\bf k}|) \approx \hat{K} (0) + 
\sum^N_{j=1} a_{\alpha_j} |{\bf k}|^{\alpha_j} ,
\ee
where the frequency dispersion is neglected, and thus $\hat{K} (0)$, 
and $a_{\alpha_j}$ ($j=1,...,N$) do not depend on the frequency $\omega$.

If $\alpha_j=j$ for all $j \in \mathbb{N}$, we can use the usual Taylor's formula. 
In this case we have a well-known weak spatial dispersion.
In general, we should use a fractional generalization of 
the Taylor's series \cite{SKM1,SKM2,Hardy,Arm1,Arm2,TRB,OdbSh}.
The orders of the fractional Taylor series approximation 
should be correlated with the orders of power-laws of weak spatial dispersions, 
which are experimentally determined.
In this case the fractional Taylor series approximation of 
$\hat{K} (|{\bf k}|)$ will be the best approximation.

We consider models of lattices with weak spatial dispersion and 
their continuous limits by using the methods suggested in \cite{JPA2006,JMP2006}.
In this limit we obtain the fractional Laplacian in the Riesz's form 
since the inverse Fourier's integral transform of $|{\bf k}|^{\alpha}$
gives the fractional Laplacian $(-\Delta)^{\alpha/2}$.
Note that some equivalence of lattice-type networks 
with long-range interactions and continuum models
for elasticity theory is considered in \cite{PZ}, 
where  the Marchaud fractional derivatives are used.


For $\alpha > 0$ and $x \in \mathbb{R}^n$, 
the fractional Laplacian in the Riesz's form
is defined in terms of the Fourier transform ${\cal F}$ by
\be \label{RFD-1}
((-\Delta)^{\alpha/2} f)(x)= {\cal F}^{-1} \Bigl( |k|^{\alpha} ({\cal F} f)(k) \Bigr) .
\ee
For $\alpha >0$, the fractional Laplacian in the Riesz's form
usually is defined in the form of the hyper-singular integral by
\[ ((-\Delta)^{\alpha/2}f)(x) =\frac{1}{d_n(m,\alpha)} \int_{\mathbb{R}^n} 
\frac{1}{|z|^{\alpha+n}} (\Delta^m_z f)(z) \, dz , \] 
where $m> \alpha$, and $(\Delta^m_z f)(z)$ is a finite difference of
order $m$ of a function $f(x)$ with a vector step $z \in \mathbb{R}^n$
and centered at the point $x \in \mathbb{R}^n$:
\[ (\Delta^m_z f)(z) =\sum^m_{k=0} (-1)^k \frac{m!}{k!(m-k)!}  \, f(x-kz) . \]
The constant $d_n(m,\alpha)$ is defined by
\[ d_n(m,\alpha)=\frac{\pi^{1+n/2} A_m(\alpha)}{2^{\alpha} 
\Gamma(1+\alpha/2) \Gamma(n/2+\alpha/2) \sin (\pi \alpha/2)} ,  \]
where
\[ A_m(\alpha)=\sum^m_{j=0} (-1)^{j-1} \frac{m!}{j!(m-j)!} \, j^{\alpha} . \]
Note that the hyper-singular integral $((-\Delta)^{\alpha/2}f)(x)$ 
does not depend on the choice of $m>\alpha$.
The Fourier transform ${\cal F}$ of the fractional Laplacian is given by 
\[ ({\cal F} (-\Delta)^{\alpha/2} f)(k) = |k|^{\alpha} ({\cal F}f)(k) . \]
This equation is valid for Lizorkin space \cite{SKM1,SKM2}
and the space $C^{\infty}(\mathbb{R}^n)$ of infinitely differentiable 
functions on $\mathbb{R}^n$ with compact support.


\section{Fractional Elasticity Equation from Lattice Model}

In the continuous limit the equations for lattices 
with weak spatial dispersion of a power-law type
gives the continuum equation for the fractional elasticity model.

{\bf Statement.} 
{\it In the continuous limit $\Delta x \to 0$, the lattice equations 
\be \label{2-54b}
M \, \frac{\partial^2 u_n(t)}{\partial t^2} = 
g  \sum_{\substack{m=-\infty \\ m \ne n}}^{+\infty} \; 
K(n-m) \; \Bigl( u_n(t) -u_m(t) \Bigr) + F (n) ,
\ee
with weak spatial dispersion of the form (\ref{approx-1})
gives the fractional continuum equation  
\be \label{2-CME}
\frac{\partial^2 u(x,t)}{\partial t^2} =
- \sum^N_{j=1} G_{\alpha_j} \, ((-\Delta)^{\alpha_j/2} u) (x,t) +
\frac{1}{\rho} \, f(x) ,
\ee
with the fractional Laplacian $(-\Delta)^{\alpha_j/2}$ of order $\alpha_j$.  
Here the variables $x$ and $\Delta x$ are dimensionless,
$f(x)=F(x)/A |\Delta x|$, $\rho= M / (A |\Delta x| )$, 
$A$ is the cross-sectional area of the medium, and 
\be \label{Gj} 
G_{\alpha_j}= \frac{g \, a_{\alpha_j} \, |\Delta x|^{\alpha_j}}{M}  \quad (j=1, . . . , N) 
\ee
are finite parameters.
\et

\bp
Let us describe the main points in the proof of this statement. 
Here we use the methods from \cite{JPA2006,JMP2006} and \cite{TarasovSpringer}.
The Fourier series transform ${\cal F}_{\Delta}$ of equation (\ref{2-54b})
gives (\ref{2-20eq}).
After division by the cross-sectional area of the medium $A$ 
and the inter-particle distance $|\Delta x|$, 
the limit $\Delta x \to 0$ for equation (\ref{2-20eq}) gives
\be \label{2-Eq-k}
\frac{\partial^2}{\partial t^2} \hat{u}(k,t) =
\sum^N_{j=1} \frac{g \, |\Delta x|^{\alpha_j}}{M} \; \hat{\mathcal{K}}_{\alpha_j, \Delta}(k) \; \hat{u}(k,t)  
+ \frac{1}{\rho} \mathcal{F}_{\Delta} \{ f (n) \}  , 
\ee
where $\rho= M/ A |\Delta x|$ is the mass density, 
$f(n) = F(n)/ A |\Delta x|$ is the force density, 
$|\Delta x|$ is the inter-particle distance, 
$A$ is the cross-sectional area of the material, and
\[ \hat{\mathcal{K}}_{\alpha_j, \Delta}(k) = - a_{\alpha_j} |k|^{\alpha_j} . \]
Here we use (\ref{approx-1}), with $G_{\alpha_j}$ ($j=1, . . . ,N$) 
are finite parameters that are defined by (\ref{Gj}). 
The expression $\hat{\mathcal{K}}_{\alpha_j, \Delta}(k)$ can be considered
as a Fourier series transform of the interaction term.
Note that $g a_{\alpha_j}  \to \infty$
for the limit $\Delta x \to 0$, if $G_{\alpha_j}$ are finite parameters.

In the limit $\Delta x \to 0$, equation (\ref{2-Eq-k}) gives
\be \label{2-Eq-k2}
\frac{\partial^2 \tilde{u}(k,t)}{\partial t^2} =
\sum^N_{j=1} G_{\alpha_j} \, \hat{\mathcal{K}}_{\alpha_j}(k) \, \tilde{u}(k,t)  
+ \frac{1}{\rho} \mathcal{F} \{ f (x) \}  , 
\ee
where $\hat{\mathcal{K}}_{\alpha_j}(k) =
\operatorname{Lim}\hat{\mathcal{K}}_{\alpha_j, \Delta}(k) 
= - a_{\alpha_j} |k|^{\alpha_j}$. 
The inverse Fourier transform of (\ref{2-Eq-k}) gives (\ref{2-CME}).
Here, we use the connection between the Riesz fractional 
Laplacian and its Fourier transform \cite{SKM1,SKM2,KST} in the form
$|k|^{\alpha_j} \longleftrightarrow (-\Delta)^{\alpha_j/2}$. \\

To illustrate this Statement we give a few examples.

{\bf Example 1.} 
If we can use the weak spatial dispersion in the form
\be \label{a2}
\hat{K} (k) \approx \hat{K} (0) + a_2 \, k^2 ,
\ee
then we obtain the well-known one-dimensional equation for elastic continuum
\be 
\frac{\partial^2 u(x,t)}{\partial t^2} = G_2 \, \Delta u(x,t) + \frac{1}{\rho} \, f(x) ,
\ee
where
\[ G_2 = \frac{g \, a_2 \, |\Delta x|^2}{M \, A} = \frac{E}{\rho} , \] 
Here $E=K |\Delta x|/A$ is Young's modulus,
$K= g a_2$ is the spring stiffness, and $\rho= M/ A |\Delta x|$ is the mass density. 
The corresponding dispersion relation is $\omega^2(k)= G_2 k^2$.

{\bf Example 2.}
If the spatial dispersion law has the form
\be \label{a2b}
\hat{K} (k) \approx \hat{K} (0) + a_2 \, k^2 + a_4 \, k^4 ,
\ee
then we derive the equation of the gradient elasticity \cite{AA2011} as
\be 
\frac{\partial^2 u(x,t)}{\partial t^2} = G_2 \, \Delta u(x,t) - 
G_4 \, \Delta^2 u(x,t)  +  \frac{1}{\rho} \, f(x) ,
\ee
where the constant of phenomenology model is associated with the lattice constants by
\[ G_4 = \frac{g \, a_4 \, |\Delta x|^4}{M \, A} = \frac{a_4 \, E \, |\Delta x|^2}{a_2 \, \rho} . \]
The correcponding dispersion relation is thus $\omega^2(k)= G_2 k^2+G_4 k^4$.
Note that the parameter $l^2$ of the gradient elasticity is related with
the coupling constants of the lattice by the equation
\be \label{L2}
l^2 =\frac{ \left| a_4 \right| \, |\Delta x|^2}{|a_2|} .
\ee
The sign of the second-gradient term is defined by 
$\operatorname{sgn} (a_4/a_2)$. 
Note that earlier it was thought that a phenomenological model of the gradient elasticity 
with a minus sign does not have the appropriate microscopic model, 
and it is thus considered one of its main weaknesses \cite{AA2011}. 
The proposed lattice model radically changes the situation.

{\bf Example 3.}
If we can use the fractional spatial dispersion law in the form
\be \label{a2b2}
\hat{K} (k) \approx \hat{K} (0) + a_2 \, k^2 + a_{\alpha} \, k^{\alpha} ,
\ee
then the fractional elasticity equation is
\be \label{fge-1}
\frac{\partial^2 u(x,t)}{\partial t^2} = G_2 \, \Delta u(x,t) - 
G_{\alpha} \, (-\Delta)^{\alpha/2} u(x,t)  +  \frac{1}{\rho} \, f(x) ,
\ee
where  
\[ G_{\alpha} = \frac{g \, a_{\alpha} \, |\Delta x|^{\alpha}}{M \, A} =
 \frac{a_{\alpha} \, E \, |\Delta x|^{\alpha-2}}{a_2 \, \rho} . \]
The correcponding dispersion relation is $\omega^2(k)= G_2 k^2+G_{\alpha} |k|^{\alpha}$.
Equation (\ref{fge-1}) define the fractional elasticity model for
the one-dimensional case ($x \in \mathbb{R}$).


\section{Solution of Fractional Elasticity Equations}


Using the same methods as above, we can derive a general model of three-dimensional lattice 
with fractional weak spatial dispersion of the form
\be  \label{peps-G}
\hat{K} ({\bf k}) = \hat{K} (0) + \sum^N_{j=1} a_{\alpha_j} \, |{\bf k}|^{\alpha_j} . 
\ee
Then the continuum equation for fractional elasticity model has the form
\be \label{2-CME-3}
\frac{\partial^2 u({\bf r},t)}{\partial t^2} =
- \sum^N_{j=1} c_j \, ((-\Delta)^{\alpha_j/2} u) ({\bf r},t) +
\frac{1}{\rho} \, f({\bf r}) ,
\ee
where we use $c_j$ as a new notation for the constants instead of $G_{\alpha_j}$ used in the one-dimensional case.
Note that ${\bf r}$ and $r=|{\bf r}|$ are dimensionless.

\subsection*{Static equation and its solution}

Let us consider the statics ($\partial u({\bf r},t) /\partial t =0$, i.e. $u({\bf r},t) = u({\bf r})$) 
in the suggested fractional elasticity model.
Then equation (\ref{2-CME-3}) gives
\be \label{FPDE-1}
\sum^N_{j=1} c_j \, ((-\Delta)^{\alpha_j/2} u) ({\bf r}) = \frac{1}{\rho} f({\bf r}) .
\ee

Equation (\ref{2-CME-3}) 
has a particular solution $u({\bf r})$ for the case $\alpha_N > 1$ and $c_N \ne 0$
(see Section 5.5.1. pages 341-344 in  \cite{KST}).  
The particular solution is represented in the form of 
the convolution of the functions $G^n_{\alpha}({\bf r})$ and $f({\bf r})$ as follow
\be \label{phi-G}
u ({\bf r})=  \frac{1}{\rho} \, \int_{\mathbb{R}^n} G^n_{\alpha} ({\bf r} - {\bf r}^{\prime}) \, 
f ({\bf r}^{\prime}) \, d^n {\bf r}^{\prime} ,
\ee
where $n=1,2,3$ and the function $G^n_{\alpha}({\bf r})$ is the Green function 
that is given by
\be \label{FGF}
G^n_{\alpha}({\bf r})= 
\int_{\mathbb{R}^n} \left( \sum^N_{j=1} c_j |{\bf k}|^{\alpha_j} \right)^{-1} \
e^{ + i ({\bf k},{\bf r}) } \, d^n {\bf k} ,
\ee
where $\alpha=(\alpha_1,...,\alpha_N)$.
The Green function (\ref{FGF}) can be simplified by using the relation (Lemma 25.1 of \cite{SKM1,SKM2}) of the form
\be \label{3-1}
\int_{\mathbb{R}^n} e^{  i ({\bf k},{\bf r}) } \, f(|{\bf k}|) \, d^n {\bf k}= 
\frac{(2 \pi)^{n/2}}{ |{\bf r}|^{(n-2)/2}} 
\int^{\infty}_0 f( \lambda) \, \lambda^{n/2} \, J_{n/2-1}(\lambda |{\bf r}|) \, d \lambda .
\ee
Here $J_{\nu}$ is the Bessel function of the first kind. 
As a result, the Fourier transform of a radial function is also a radial function.
Using relation (\ref{3-1}), the Green function (\ref{FGF}) 
can be represented (see Theorem 5.22 in \cite{KST})
in the form of an integral with respect to one parameter $\lambda$,
\be \label{G-1}
G^n_{\alpha} ({\bf r}) = \frac{|{\bf r}|^{(2-n)/2}}{(2 \pi)^{n/2}} 
\int^{\infty}_0 \left( \sum^N_{j=1} c_j \lambda^{\alpha_j} \right)^{-1} 
\lambda^{n/2} \, J_{(n-2)/2} (\lambda |{\bf r}|) \, d \lambda,
\ee
where $n=1,2,3$ and $\alpha=(\alpha_1,...,\alpha_m)$, and 
$J_{(n-2)/2}$ is the Bessel function of the first kind, which
can be represented as
$J_{1/2} (z) = \sqrt{2 / \pi z} \, \sin (z)$
for the 3-dimensional case.

\subsection*{Thomson's problem for fractional integral and gradient elasticity}

If we have the dispersion law in the form
\be \label{a2c}
\hat{K} (|{\bf k}|) \approx \hat{K} (0) + a_{\alpha} |{\bf k}|^{\alpha} + a_2 |{\bf k}|^2 ,
\ee
where $\alpha>0$, then we obtain the fractional elasticity equation 
\be \label{FPDE-4}
c_2 \Delta u ({\bf r}) - c_{\alpha} \, ((-\Delta)^{\alpha/2} u) ({\bf r}) + \frac{1}{\rho} \, f({\bf r}) =0 ,
\ee
where
\be \label{cGj} 
c_2 = \frac{E}{\rho} = \frac{g \, a_2 \, |\Delta x|^2}{M} , \quad 
c_{\alpha}= \frac{g \, a_{\alpha} \, |\Delta x|^{\alpha}}{M} .
\ee
If $\alpha=4$ we have the well-known static equation of the gradient elasticity \cite{AA2011}:
\be \label{GradEl}
c_2 \, \Delta u ({\bf r}) - c_4 \Delta^2 u ({\bf r}) + \frac{1}{\rho} \, f({\bf r}) = 0,
\ee
where
\be \label{GradEl-2} 
c_4 = \pm \, l^2 \, \frac{E}{\rho}= \frac{g \, a_4 \, |\Delta x|^4}{M} .
\ee
The second-gradient term is preceded by the sign that is defined by
$\operatorname{sgn} (g \, a_4)$, where $g \, a_2>0$.

Equation (\ref{FPDE-4}) with $n=3$ has the particular solution \cite{KST} of the form
\be \label{phi-G4}
u({\bf r})=  \frac{1}{\rho} \, \int_{\mathbb{R}^3} 
G^3_{\alpha} ({\bf r} - {\bf r}^{\prime}) \, 
f({\bf r}^{\prime}) \, d^3 {\bf r}^{\prime},
\ee
where the Green type function is given by
\be \label{G-4}
G^3_{\alpha} ({\bf r}) =\frac{|{\bf r}|^{-1/2}}{(2 \pi)^{3/2}} 
\int^{\infty}_0 \left( c_{\alpha} \lambda^{\alpha}+ c_2 |\lambda|^2 \right)^{-1} 
\lambda^{3/2} \, J_{1/2} (\lambda |{\bf r}|) \, d \lambda .
\ee
Here $J_{1/2}$ is the Bessel function of the first kind.


Let us consider W. Thomson (1848) problem \cite{LL}
for the fractional elasticity model described by equation (\ref{FPDE-4}). 
This problem implies that we should determine the deformation 
of an infinite elastic continuum,
when a force is applied to a small region in it.
If we consider the deformation at distances $|{\bf r}|$, 
which are larger compared to the size of the region,
we can assume that the force is applied at a point, i.e.
\be \label{deltaf}
f({\bf r}) = f_0 \, \delta ({\bf r}) = f_0 \, \delta (x) \delta (y) \delta (z)  . 
\ee
Then the displacement field $u ({\bf r})$ of fractional elasticity 
has a simple form of the particular solution  
that is proportional to the Green's function
\be \label{phi-Gb}
u ({\bf r}) = \frac{f_0}{\rho} \, G^n_{\alpha} ({\bf r}) .
\ee 
As a result the displacement field for the force that is applied at a point (\ref{deltaf}) has the form
\be \label{Pot-2}
u ({\bf r}) = \frac{1}{2 \pi^2} \frac{f_0}{ \rho \, |{\bf r}|} \, 
\int^{\infty}_0 \frac{ \lambda \, \sin (\lambda |{\bf r}|)}{ 
c_{\alpha} \lambda^{\alpha}+ c_2 \lambda^2  } \, d \lambda .
\ee


We can distinguish the following two cases:
(1) Weak power-law spatial dispersion with $\alpha < 2$;
(2) Weak power-law spatial dispersion with $\alpha > 2$.
This is due to the fact that in nonlocal elasticity theory 
usually distinguish the following two cases:
(1) Fractional integral elasticity ($\alpha<2$); 
(2) Fractional gradient elasticity ($\alpha > 2$). 

For the fractional integral elasticity, the order of the fractional Laplacian 
is less than the order of the term related to Hooke's law. 
For the fractional gradient elasticity, the order of the fractional Laplacian 
is greater then the order of the Hooke's term.


\subsection*{Fractional integral elasticity model}

The fractional integral elasticity model is described by equation 
(\ref{FPDE-4}) with $\alpha<2$ of the form 
\be \label{FPDE-4-2b1}
c_2 \Delta u ({\bf r}) - c_{\alpha} ((-\Delta)^{\alpha/2} u) ({\bf r}) + \frac{1}{\rho} \, f({\bf r}) = 0 ,
\quad (0<\alpha<2) .
\ee
The order of the fractional Laplacian $(-\Delta)^{\alpha/2}$ is less 
than the order of the first term related to the usual Hooke's law. 
Note that the continuum equation (\ref{FPDE-4-2b1}) of fractional integral elasticity
is derived from the lattice equations with weak spatial dispersion 
in the form (\ref{a2c}) with $\alpha< 2$.
The particular solution of equation (\ref{FPDE-4-2b1})
for the force that is applied at a point (\ref{deltaf}) is the displacement field 
\be \label{Pot-2-2b}
u ({\bf r}) = \frac{f_0}{2 \pi^2 \rho \, |{\bf r}|} \, 
\int^{\infty}_0 \frac{ \lambda \, \sin (\lambda |{\bf r}|)}{ 
c_2 \lambda^2+ c_{\alpha} \lambda^{\alpha}  } \, d \lambda \quad (\alpha <2).
\ee
Using Section 2.3.1 in the book \cite{BE}, 
we can obtain the asymptotic behavior for (\ref{Pot-2-2b}) for 
\be
u ({\bf r})  \approx
 \frac{ C_0(\alpha) }{|{\bf r}|^{3-\alpha}} + \sum^{\infty}_{k=1}  \frac{ C_k(\alpha) }{|{\bf r}|^{(2-\alpha)(k+1)+1}} 
\quad (|{\bf r}| \to \infty),
\ee
where 
\be
C_0(\alpha)= \frac{f_0}{2\pi^2 \, \rho \, c_{\alpha}} \, \Gamma(2-\alpha) \, \sin \left( \frac{\pi}{2}\alpha \right) ,
\ee
\be
C_k (\alpha) = -\frac{f_0 c^k_2}{2 \pi^2 \,\rho \, c^{k+1}_{\alpha}} \int^{\infty}_0  z^{(2-\alpha)(k+1)-1} \, \sin(z) \, dz .
\ee
As a result, the displacement field for the force that is applied at a point
in the continuum with this type of fractional weak spatial dispersion is given by
$u ({\bf r}) \ \approx \ \, C_0 (\alpha)/ |{\bf r}|^{3-\alpha}$, where $0< \alpha<2$,
on the long distance $|{\bf r}| \gg 1$. 
The asymptotic behavior $|{\bf r}| \to 0$ 
for the fractional integral elasticity does not depend on the parameter $\alpha$.


\subsection*{Fractional gradient elasticity model}

The fractional gradient elasticity model is described by the equation 
\be \label{FPDE-4-2b2}
c_2 \Delta u ({\bf r}) - c_{\alpha} ((-\Delta)^{\alpha/2} u) ({\bf r}) +  \frac{1}{\rho} \, f({\bf r}) = 0 ,
\quad (\alpha > 2) .
\ee
This may be derived from the lattice model with the fractional weak spatial dispersion 
in the form (\ref{a2c}) with $\alpha > 2$.
The order of the fractional Laplacian $(-\Delta)^{\alpha/2}$ 
is greater than the order of the first term related to the Hooke's law.
If $\alpha =4$ equation (\ref{FPDE-4-2b2}) become the equation (\ref{GradEl}). 
Therefore the case $3<\alpha<5$ can be considered as close as possible ($\alpha \ \approx \ 4$)
to the usual gradient elasticity (\ref{GradEl}).
The continuum equation (\ref{FPDE-4-2b2}) of fractional gradient elasticity
is derived from the lattice equations with weak spatial dispersion 
in the form (\ref{a2c}) with $\alpha > 2$.

The particular solution of equation (\ref{FPDE-4-2b2})
for the force that is applied at a point (\ref{deltaf}) is the displacement field 
\be \label{Pot-2-2c}
u ({\bf r}) = \frac{f_0}{2 \pi^2 \rho \, |{\bf r}|} \, 
\int^{\infty}_0 \frac{ \lambda \, \sin (\lambda |{\bf r}|)}{ 
c_2 \lambda^2+ c_{\alpha} \lambda^{\alpha}  } \, d \lambda \quad (\alpha >2).
\ee
The asymptotic behavior $|{\bf r}| \to \infty$ 
for the fractional gradient elasticity does not depend on the parameter $\alpha$.
The asymptotic behavior of the displacement field $u ({\bf r})$ for $|{\bf r}| \to 0$ is given by
\be \label{Cab-2}
u ({\bf r}) \ \approx \ 
\frac{f_0  \, \Gamma((3-\alpha)/2)}{2^{\alpha} \, \pi^2 \sqrt{\pi} \, \rho \, c_{\alpha} \, \Gamma(\alpha/2)} \,
\cdot \,  |{\bf r}|^{\alpha-3} , \quad (2<\alpha<3),
\ee
\be \label{Cab-3}
u ({\bf r}) \ \approx \ 
\frac{f_0}{2 \pi \, \alpha \, \rho \, c^{1-3/\alpha}_2 \, c^{3/\alpha}_{\alpha} \, \sin (3 \pi / \alpha)}
 , \quad (\alpha>3) .
\ee
Note that the asymptotic behavior for $2 < \alpha <3$ does not depend on $c_2$.
The displacement field $u ({\bf r})$ of the short distance 
is determined only by term with $(-\Delta)^{\alpha/2}$ ($\alpha>2$) 
which can thus be considered as a fractional non-locality of the gradient type. 

\section{Conclusion}

We demonstrate the close relation between 
the discrete microstructure of a lattice with weak spatial dispersion
of a power-law type and the fractional integral and gradient elasticity 
models. 
We prove that fractional elasticity models can be directly derived from the lattice models with
power-law spatial dispersion.
It has been shown that a characteristic feature of the behavior 
of a fractional non-local continuum is the spatial power-tails
of non-integer orders.
We assume that fractional elastic media (and plasma-like media with power-law spatial dispersion \cite{AP2013})
should demonstrate a universal behavior in space by analogy 
with the universal behavior of low-loss dielectrics in time \cite{Jo1,Jo2,Jo3,JPCM2008-1,JPCM2008-2}.
Note that the universal behavior in time can be connected with 
the interaction between the particles and the environment \cite{CEJP}.




\end{document}